\documentclass{svproc}
\usepackage{amsmath}
\usepackage{amssymb}

\usepackage{graphicx}
\usepackage{subcaption}
\usepackage[caption = false]{subfig}
\usepackage{array}
\usepackage{url}

\begin{document}
\mainmatter              
\title{Numerical Simulations for Time-Fractional Black Scholes Equations}
\titlerunning{Time-Fractional Black Scholes}  
%
\author{Neetu Garg\inst{1} \and  A.S.V. Ravi Kanth\inst{2}}
\authorrunning{Neetu Garg et al.} 
%
\tocauthor{Neetu Garg and A.S.V. Ravi Kanth}
\institute{National Institute of Technology Calicut,\\
	Kerala, India\\
\email{neetu@nitc.ac.in}
\and
National Institute of Technology Kurukshetra,\\
Haryana, India\\
\email{asvravikanth@yahoo.com} }

\maketitle              

\begin{abstract}
This paper implements an efficient numerical algorithm for the time-fractional Black-Scholes model governing European options. The proposed method comprises the Crank-Nicolson approach to discretize time variable and exponential B-spline approximation for space variable. The implemented method is unconditionally stable. We present few numerical examples to confirm the theory. Numerical simulations with comparisons exhibit the supremacy of the proposed approach.
\keywords{Caputo fractional derivative, Exponential B-spline, Black-Scholes equation, Stability}
\end{abstract}
\section{Introduction}
Option pricing is a significantly crucial concept in the financial market. Black \cite{Black} and Merton \cite{Merton} came up with the idea of the Black-Scholes equation for option pricing. There has been an enormous research activity in financial mathematics after the publication of the Black-Scholes model. Many researchers developed a huge interest in studying both theoretical and practical aspects of European options. 

Recently, fractional calculus has gained huge attention as it provides an excellent instrument to characterize the memory phenomena due to the non-locality of fractional derivative \cite{Podlubny}. The time-fractional Black-Scholes model (TFBSM) has received enormous popularity because of its ability to capture significant jumps during small time durations. At first, the European call option was priced by Wyss \cite{Wyss} via fractional model. In the literature, several analytical and numerical methods had been reported including, finite difference method, implicit finite difference method, compact difference scheme, integral discretization method, meshless method, residual power series method, moving least square method, radial basis functions, quintic B-spline method, finite element method \cite{Song,HZhang,Cen,Haq,Golba,Roul,Ram1,Ram2,Ram3}.
This paper focuses on the following TFBSM:
\begin{eqnarray}
	\label{eq1}
	\frac{\partial ^\mu \mathcal{W}(\zeta,t)}{\partial t^\mu } +\frac{1}{2}\sigma ^2\zeta^2\frac{\partial ^2 \mathcal{W}(\zeta,t)}{\partial \zeta^2 }+ (\mathfrak{r}-\mathfrak{D}) \zeta \frac{\partial \mathcal{W}(\zeta,t)}{\partial \zeta }=\mathfrak{r}\mathcal{W}(\zeta,t),\\
	(\zeta,t)\in (0,\infty)\times(0,\mathcal{T}), ~0< \mu \leq 1,\nonumber
\end{eqnarray} along with conditions
\begin{equation}
	\label{eq2}
	\mathcal{W}(\zeta,\mathcal{T}) = \mathcal{W}(\zeta),\\
	\mathcal{W}(0,t) = h_1(t), \mathcal{W}(\infty,t) = h_2(t),
\end{equation}
where $\mathcal{W}(\zeta,t)$ is the European option price with stock price $\zeta$ and current time $t$, $\mathfrak{r}$ is the risk-free interest rate, $\mathfrak{D}$ is the dividend rate, $\sigma$ is volatility and $\mathcal{T}$ is the expiry time. Here, the modified Riemann-Liouville fractional derivative $\frac{\partial ^\mu \mathcal{W}(\zeta,t)}{\partial t^\mu}$ is given as
\begin{equation}
	\label{eq4}
	\frac{\partial ^\mu \mathcal{W}(\zeta,t)}{\partial t^\mu }\nonumber =\left\{\begin{matrix}
		{\Gamma(1-\mu)}^{-1}\frac{d}{dt} \int_{t}^{\mathcal{T}}\frac{\mathcal{W}(\zeta,\alpha)-\mathcal{W}(\zeta,\mathcal{T})}{(\alpha-t)^\mu }d\alpha, & 0<\mu <1\\
		\frac{\partial \mathcal{W}(\zeta,t)}{\partial t},& \mu=1
	\end{matrix}\right.
\end{equation}
Using the transformation $t=\mathcal{T}-\tau$ ($0<\mu <1$), we obtain
\begin{eqnarray}
	\label{eq5}
	\frac{\partial ^\mu \mathcal{W}(\zeta,t)}{\partial t^\mu }
	=-{\Gamma(1-\mu)}^{-1}\frac{d}{d\tau} \int_{0}^{\tau}\frac{\mathcal{W}(\zeta, \mathcal{T}-\beta )-\mathcal{W}(\zeta,\mathcal{T})}{(\tau-\beta)^\mu }d\beta.
\end{eqnarray}
Assuming $y=\ln \zeta$ and $u(y,\tau)= \mathcal{W}(e^y, \mathcal{T}-\tau)$, the model \eqref{eq1}-\eqref{eq2} turns into
\begin{equation}
	\label{eq6}
	\left\{\begin{matrix}
		_{0}\textrm{D}_{\tau}^{\mu}u(y,\tau)= \frac{\sigma^2}{2}\frac{\partial^2 u(y,\tau)}{\partial y^2}+ \left (\mathfrak{r}- \frac{\sigma^2}{2}-\mathfrak{D} \right )\frac{\partial  u(y,\tau)}{\partial y}- \mathfrak{r}u(y,\tau) \\
		u(-\infty, \tau)= h_1(\tau),~ u(\infty, \tau) = h_2(\tau), \\
		u(y,0)= u_0(y),
	\end{matrix}\right.
\end{equation} where $_{0}\textrm{D}_{\tau}^{\mu}u(y,\tau)$ denotes
\begin{align}
	\label{eq7}
	_{0}\textrm{D}_{\tau}^{\mu}u(y,\tau)={\Gamma(1-\mu)}^{-1}\frac{d}{d\tau} \int_{0}^{\tau}\frac{u(y, \beta )-u(y,0)}{(\tau-\beta)^\mu }d\beta, 0<\mu<1.
\end{align}
For solving \eqref{eq6} numerically, the unbounded domain is truncated to a finite domain $(y_a,y_b)$, for details see \cite{Ram4}. Thus we have the dimensionless model as \cite{frac}
\begin{eqnarray}
	\label{eq8}
	\left\{\begin{matrix}
		\textrm{D}_{\tau}^{\mu}u(y,\tau)= \kappa_1\frac{\partial^2 u(y,\tau)}{\partial y^2}+ \kappa_2\frac{\partial  u(y,\tau)}{\partial y}- \kappa_3 u(y,\tau)+ g(y,\tau),\\
		u(y_a, \tau)= h_1(\tau),~ u(y_b, \tau) = h_2(\tau),\\
		u(y,0)= u_0(y), y_a < y< y_b,
	\end{matrix}\right.
\end{eqnarray} where $\kappa_1= \frac{\sigma^2}{2}>0$, $\kappa_2=\mathfrak{r}-\mathfrak{D}-\kappa_1$, $\kappa_3=\mathfrak{r}>0$. Here we add the smooth force term $g(y,t)$ for validation purpose in section 5.

%
In this work, we present an efficient technique for the TFBSM. We first employ a Crank-Nicolson approach for discretizing time variable \cite{Karatay} and then exponential B-spline functions are used for approximating the resulting equation. We validate the proposed method via numerical experiments. To exhibit the efficiency of the proposed algorithm, we also carry out comparisons with the existing results.
It is noteworthy that the exponential B-spline method is one of the most robust numerical methods based on piecewise non-polynomial basis functions of class $\mathcal{C}^2$ with compact support \cite{McCartin}. In contrast to the finite element method, it saves us from the computation of quadratures. In literature, the exponential B-spline method has contributed in solving a wide range of problems \cite{ravi,ravi2}.
The article is planned as follows. Section 2 develops the exponential B-spline approximation to TFBSM. Section 3 discusses stability analysis. In section 4, we study numerical results to assess the validity and accuracy of the proposed scheme. Finally, the main conclusions are summarized.
\section{Methodology}
In this section, we derive a numerical method by comprising a Crank-Nicolson approach for time variable and exponential B-spline approximation for the space variable.
To start, we first partition the solution domain $[y_a,y_b] \times [0,\mathcal{T}]$ as
\begin{eqnarray*}
	y_j= j\Delta y~ (0\leq j \leq J),~\tau_n = n\Delta \tau ~(0\leq n \leq \mathcal{N})
\end{eqnarray*} with space and time steps $\Delta y=\frac{y_b-y_a}{J}$ and $\Delta \tau =\frac{\mathcal{T}}{\mathcal{N}}$, respectively.\\
Let $u(y,\tau)\in \mathcal{C}^{(1)}$ about time, the modified Riemann-Liouville derivative
\begin{align}\label{eq9}
	&_{0}\textrm{D}_{\tau}^{\mu}u(y,\tau)={\Gamma(1-\mu)}^{-1} \int_{0}^{\tau}\frac{du(y, \beta )}{d\beta}(\tau-\beta)^{-\mu}d\beta =_{0}^C\textrm{D}_{\tau}^{\mu}u(y,\tau)
\end{align}
Then, the Caputo derivative $_{0}^C\textrm{D}_{\tau}^{\mu}u$ at the node $(y, \tau_{n+\frac{1}{2}})$ can be discretized as \cite{Karatay}:
\begin{align}
	\label{eq10}
	&_{0}^C\textrm{D}_{\tau}^{\mu}u(y,\tau_{n+\frac{1}{2}})
	= \varpi \left(u^{n+1}(y)-u^n(y)\right)+ \nu_1 u^n(y)- \nu_n u^0(y)\nonumber\\
	&+ \sum_{q=1}^{n-1}(\nu_{n-q+1}-\nu_{n-q})u^q(y)
	+ O( \Delta \tau^{2-\mu}),~0< \mu < 1
\end{align}where
\begin{align*}
	&u^{n+1}(y)=u(y,\tau_{n+1}), \varpi= {2^{\mu-1 }\Delta \tau ^{-\mu}}{\Gamma(2-\mu )}^{-1},\nonumber\\
	&\nu_i= {\Delta \tau ^\mu\Gamma(2-\mu ) }^{-1}\left((i+ 0.5)^{1-\mu }-(i- 0.5)^{1-\mu }\right), (1\leq i \leq n).
\end{align*}\\
Employing the Crank-Nicolson approach to \eqref{eq8} at $(y,\tau_{n+\frac{1}{2}})$ and substituting \eqref{eq10} yields
\begin{align}
	\label{eq12}
	&-\kappa_1 \frac{\partial^2 \mathcal{U}^{n+1}(y)}{\partial y^2}-\kappa_2 \frac{\partial \mathcal{U}^{n+1}(y)}{\partial y}+ (2\varpi+\kappa_3) \mathcal{U}^{n+1}(y)\nonumber\\
	&= \kappa_1 \frac{\partial^2 \mathcal{U}^{n}(y)}{\partial y^2}+\kappa_2 \frac{\partial \mathcal{U}^{n}(y)}{\partial y}+ (2\varpi-\kappa_3) \mathcal{U}^n(y)+ 2 g^{n+\frac{1}{2}}\nonumber\\
	&+ 2\left (\nu_n \mathcal{U}^0(y) -\nu_1 \mathcal{U}^n(y)+ \sum_{q=1}^{n-1}(\nu_{n-q}-\nu_{n-q+1})\mathcal{U}^q(y)\right ).
\end{align} where $g^{n+\frac{1}{2}}= g(y, \tau_{n+\frac{1}{2}})$ for $n=0,1,...,\mathcal{N}-1$ and $\mathcal{U}^n(y)$ as approximate solution of $u^n(y)$.\\
Next, we discretize the equation \eqref{eq12} using the exponential B-spline functions $E\mathfrak{B}_j(y)$ ($j=-1,0,...,J+1$) defined in \cite{McCartin}.
Clearly, the basis functions $E\mathfrak{B}_j(y)$ are twice continuously differentiable and vanish outside $[y_{j-2},y_{j+2}]$.
We approximate in space via B-spline functions as \cite{McCartin}:
\begin{equation}
	\label{eq14}
	\mathcal{U}(y,\tau)= \sum_{j=-1}^{J+1}\delta _j(\tau)E\mathfrak{B}_j(y),
\end{equation} where $\delta _j(\tau)$ are undetermined coefficients to be computed. Using \eqref{eq14}, $E\mathfrak{B}_j(y)$ and its first two derivatives at the nodes $y_j$'s~ $(j=0,1,...,J)$ can be tabulated as follows:
\begin{table}[h]
	\centering
	\caption{Values of exponential B-splines}
	\label{tab:1}	
		\begin{tabular}{llllll}
			\hline
			& $y_{j-2}$ & $y_{j-1}$ & $y_{j}$ & $y_{j+1}$ & $y_{j+2}$ \\
			\hline
			$E\mathfrak{B}_j(y)$ & 0 & $\gamma_1$ & 1 & $\gamma_1$ & 0 \\
			
			$E\mathfrak{B}_j^{'}(y)$ & 0 & $\gamma_2$ & 0 & -$\gamma_2$ & 0 \\
			$E\mathfrak{B}_j^{''}(y)$ & 0 & $\gamma_3$ & -2$\gamma_3$ & $\gamma_3$ & 0 \\
			\hline
	\end{tabular}
\end{table}\\
where
\begin{equation*}
	\gamma_1 = \frac{s-\mathfrak{p} \Delta y}{2(\mathfrak{p} \Delta yc-s)},
	\gamma_2 = \frac{\mathfrak{p}(1-c)}{2(\mathfrak{p} \Delta yc-s)},
	\gamma_3 = \frac{\mathfrak{p} ^2s}{2(\mathfrak{p} \Delta yc-s)}.
\end{equation*} Here  $s=\sinh(\mathfrak{p} \Delta y), c=\cosh(\mathfrak{p} \Delta y)$ and $\mathfrak{p}$ is a non-negative parameter.
Using \eqref{eq14} and Table \ref{tab:1} in \eqref{eq12} at $y_j$'s gives
\begin{eqnarray}
	\label{eq15}
	\Im_1 \delta _{j-1}^{n+1}+ \Im_2 \delta _j^{n+1}+ \Im_3 \delta _{j+1}^{n+1}= \varphi_j ^n,~0\leq j \leq J,~0 \leq n \leq \mathcal{N}-1,
\end{eqnarray}
where
\begin{align*}
	&\varphi_j ^n= \Im_4 \delta _{j-1}^n+ \Im_5 \delta _j^n+ \Im_6 \delta _{j+1}^n \nonumber\\
	&+ 2 \left ( \nu_{n} \mathcal{G}_j^0- \nu_{1} \mathcal{G}_j^n-\sum_{q=1}^{n-1}(\nu_{n-q+1}-\nu_{n-q}) \mathcal{G}_j^q \right )+ 2g_j^{n+\frac{1}{2}}, \nonumber\\
	&\Im_1= \gamma_1(2 \varpi+\kappa_3)- \gamma_2 \kappa_2-\gamma_3 \kappa_1,\nonumber\\
	&\Im_2= 2 \varpi+\kappa_3+ 2 \gamma_3 \kappa_1, \nonumber\\
	&\Im_3= \gamma_1(2 \varpi+\kappa_3)+\gamma_2 \kappa_2-\gamma_3 \kappa_1,\nonumber\\
	&\Im_4= \gamma_1(2 \varpi-\kappa_3)+ \gamma_2 \kappa_2+\gamma_3 \kappa_1,\nonumber\\
	&\Im_5= 2 \varpi-\kappa_3- 2 \gamma_3 \kappa_1,\nonumber\\
	&\Im_6= \gamma_1(2 \varpi-\kappa_3)- \gamma_2 \kappa_2+\gamma_3 \kappa_1\nonumber\\
	&\mathcal{G}_j^m= \gamma_1 \delta _{j-1}^m+ \delta _j^m+ \gamma_1\delta _{j+1}^m, m=0,1,...,n.\nonumber
\end{align*}
To make system \eqref{eq15} solvable, we need two more equations in the form of following discretized boundary conditions
\begin{align}
	\label{eq16}
	& \gamma_1\delta _{-1}^{n+1}+ \delta _0^{n+1}+\gamma_1 \delta _{1}^{n+1} = h_1^{n+1}, \nonumber\\
	& \gamma_1\delta _{J-1}^{n+1}+ \delta _J^{n+1}+\gamma_1 \delta _{J+1}^{n+1} = h_2^{n+1}.
\end{align}
Eliminating $\delta _{-1}$ and $ \delta _{J+1}$ from \eqref{eq16} and together with \eqref{eq15}, we get the system of $J+1$ constrains in $J+1$ variables which can be solved easily.
\section{Stability Analysis}
\begin{theorem}
	\label{thm1}
	The proposed scheme \eqref{eq15} is stable unconditionally.
\end{theorem}
\textbf{Remark} The proof of the above theorem is on the similar lines as in \cite{ravi,ravi2}.
\section{Numerical results and discussion}\label{sec:4}
\begin{figure}[!htbp]
	\begin{subfigure}[b]{0.5\textwidth}
		\includegraphics[width=2.6in]{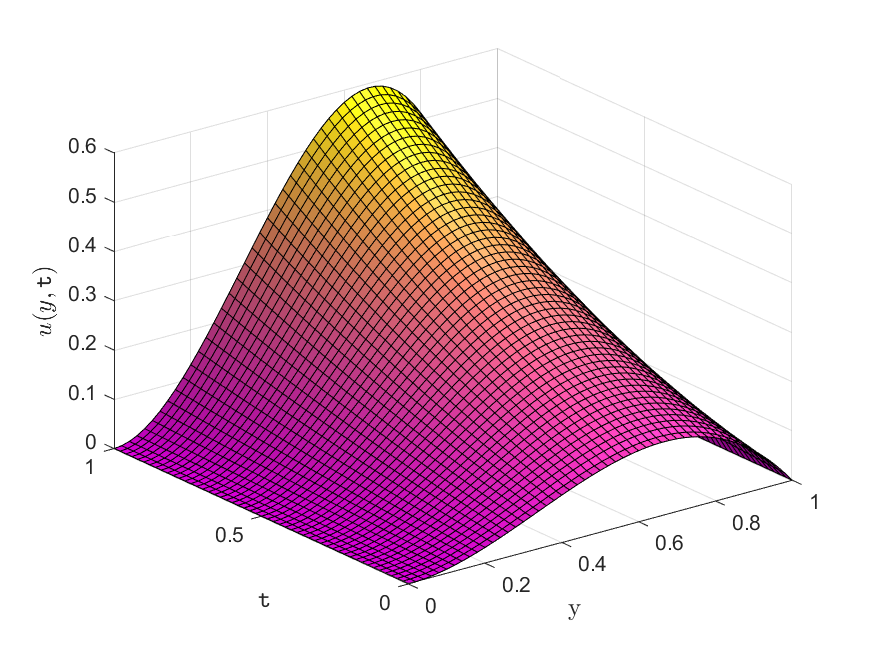}
		\subcaption{Numerical solution}
		\label{fig1a}
	\end{subfigure}
	\begin{subfigure}[b]{0.5\textwidth}
		\includegraphics[width=2.6in]{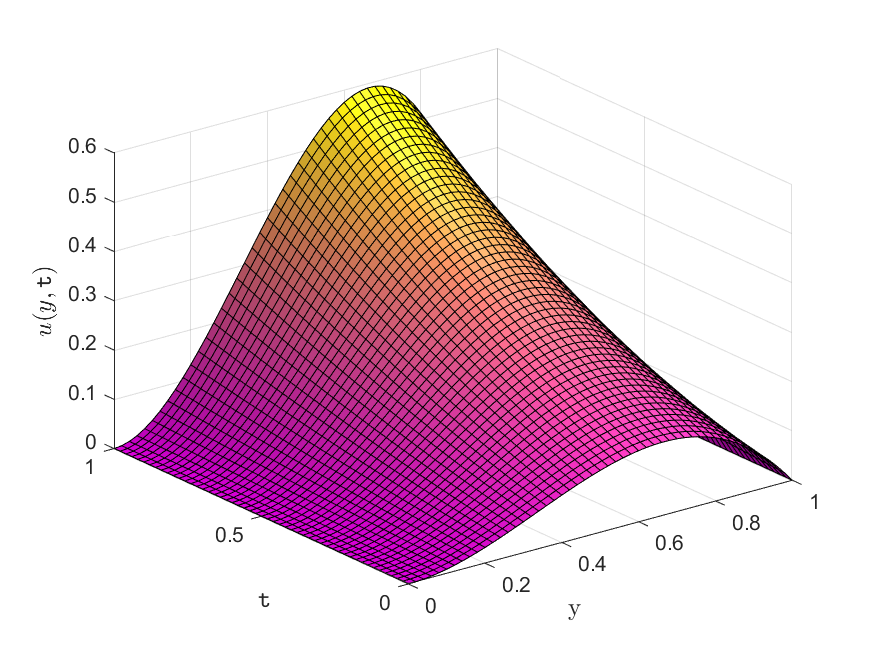}
		\subcaption{Exact solution}
		\label{fig1b}
	\end{subfigure}
	\caption{Solution profiles with $\mathfrak{p}=1$, $\mathcal{N}=J=50$ and $\mu=0.5$}
	\label{fig1}
\end{figure}

\begin{figure}[!htbp]
	\begin{subfigure}[b]{0.5\textwidth}
		\includegraphics[width=2.6in]{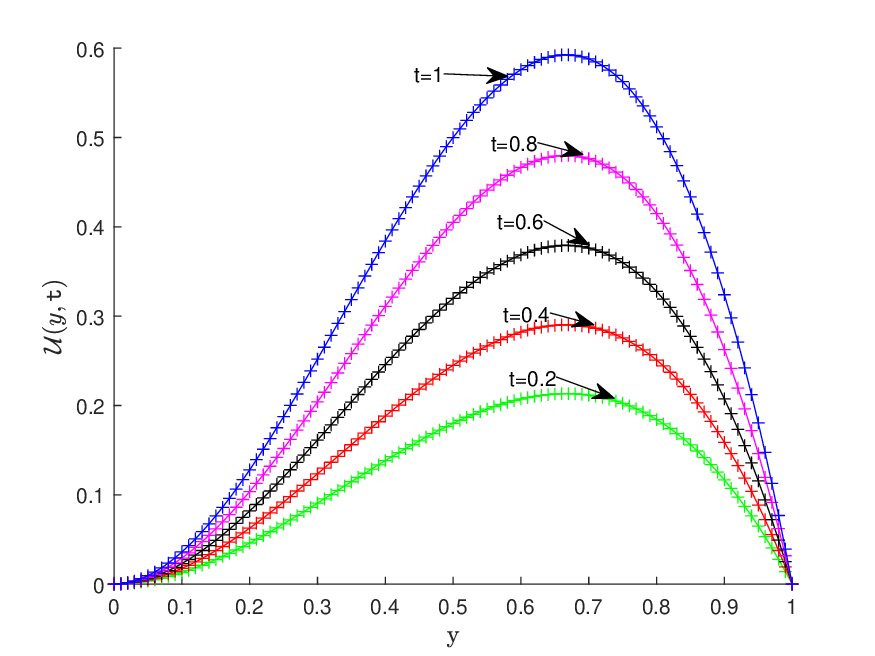}
		\caption{$\mu=0.5$}
		\label{fig2a}
	\end{subfigure}
	\begin{subfigure}[b]{0.5\textwidth}
		\includegraphics[width=2.6in]{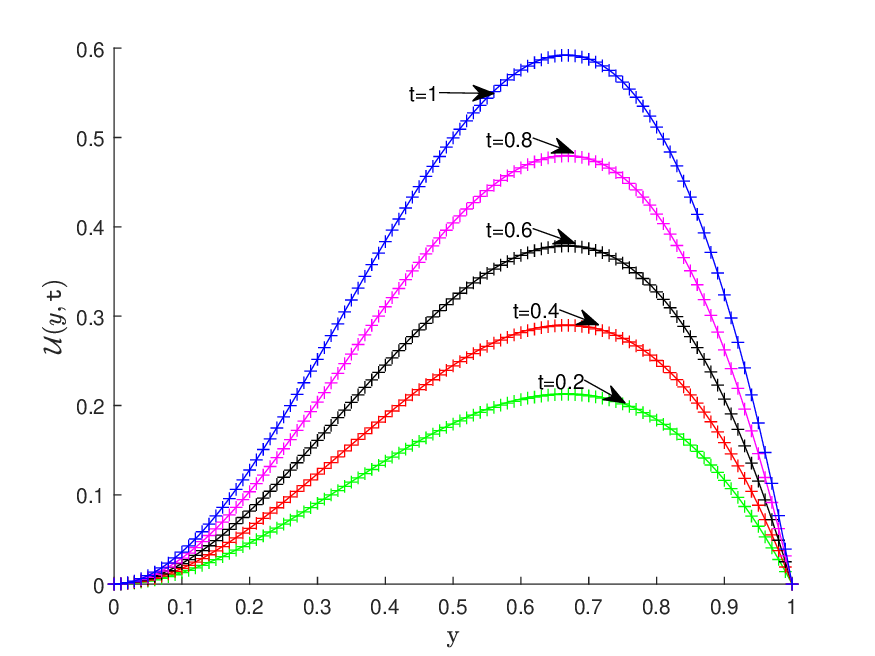}
		\caption{$\mu=0.9$}
		\label{fig2b}
	\end{subfigure}
	\caption{Numerical solution (+) and exact solution (solid) at different time levels with $\mathfrak{p}=1$ and $\mathcal{N}=J=100$}
	\label{fig2}
\end{figure}

Consider the following model (c.f. \cite{Golba}):
\begin{align*}
	&_{0}\textrm{D}_{\tau}^\mu u(y,\tau)=\kappa_1 \frac{\partial ^2u(y,\tau)}{\partial y^2}+ \kappa_2 \frac{\partial u(y,\tau)}{\partial y} -\kappa_3 u(y,\tau)+ g(y,\tau),\\
	&u(0,\tau)= u(1,\tau)=0,\\
	&u(y,0)= y^2(1-y)
\end{align*} where
\begin{align*}
	g(y, \tau)=\left ( \frac{2\tau^{2-\mu }}{\Gamma(3-\mu)} +\frac{2\tau^{1-\mu }}{\Gamma(2-\mu)}\right )y^2(1-y)-(\tau+1)^2\left [ \kappa_1(2-6y)+\kappa_2(2y-3y^2)-\kappa_3y^2(1-y) \right ].
\end{align*}The exact solution is $u(y,\tau)= (\tau+1)^2y^2(1-y)$. We choose parameters as $\mathfrak{r}=0.05, \sigma=0.25, \mathfrak{D}=0, \kappa_1=0.5 \sigma^2, \kappa_2= \mathfrak{r}-a, \kappa_3=\mathfrak{r} $ and $\mathcal{T}=1$.\\
The computed errors and corresponding rate of convergence are listed in tables \ref{tab:2} and \ref{tab:3}. It is clearly seen from these tables that the proposed method is $2-\mu$ order accurate in time and obtains second order accuracy in space for different fractional orders. Table \ref{tab:4}-\ref{tab:6} yields comparison results that are in accordance with the results in \cite{Golba,Roul,De}.
Figure \ref{fig1} plot 3-D graphs of numerical and exact solution for $\mu=0.5$. This figure clearly indicates the close proximity between the numerical and exact solutions. Figure \ref{fig2} portrays exact and numerical solution profiles at different times for $\mu=0.5$ and $0.9$. This figure presents how time affects the option pricing. 

\begin{table}[!htbp]
	\centering
	\caption{}{$\mathfrak{L}_2$ error estimate with $ \mathfrak{p}=0.01$ and $J=200$}
	\label{tab:2}
	\begin{tabular}{lllllllll}
		\hline
		\multicolumn{1}{c}{$\mathcal{N}$ } & \multicolumn{2}{c}{$\mu=0.9$}& \multicolumn{2}{c}{$\mu=0.7$}& \multicolumn{2}{c}{$\mu=0.5$}& \multicolumn{2}{c}{$\mu=0.3$}\\
		\hline
		20 & 2.1966e-03 & $-$ & 8.8682e-04 & $-$  & 3.1351e-04 & $-$  & 9.2986e-05 & $-$  \\
		40 & 1.0204e-03 & 1.1061 & 3.5957e-04 & 1.3024 & 1.1134e-04 & 1.4935 & 2.9166e-05 & 1.6727 \\
		80 & 4.7504e-04 & 1.1030 & 1.4594e-04 & 1.3009 & 3.9499e-05 & 1.4951 & 9.1103e-06 & 1.6787 \\
		160 & 2.2138e-04 & 1.1015 & 5.9252e-05 & 1.3004 & 1.3997e-05 & 1.4967 & 2.8330e-06 & 1.6852 \\
		320 & 1.0322e-04 & 1.1008 & 2.4059e-05 & 1.3003 & 4.9530e-06 & 1.4987 & 8.7453e-07 & 1.6958 \\
		640 & 4.8138e-05 & 1.1005 & 9.7665e-06 & 1.3007 & 1.7482e-06 & 1.5024 & 2.6513e-07 & 1.7218 \\
		1280 & 2.2450e-05 & 1.1005 & 3.9620e-06 & 1.3016 & 6.1329e-07 & 1.5112 & 7.6072e-08 & 1.8013 \\
		\hline
	\end{tabular}
\end{table}
\begin{table}[!htbp]
	\centering
	\caption{}{$\mathfrak{L}_2$ error estimate with $ \mathfrak{p}=1$ and $ \Delta \tau={ \Delta y}^2$}
	\label{tab:3}
	\begin{tabular}{lllllll}
		\hline
		\multicolumn{1}{c}{$J$ } & \multicolumn{2}{c}{$\mu=0.75$}& \multicolumn{2}{c}{$\mu=0.5$}& \multicolumn{2}{c}{$\mu=0.25$}\\
		\hline
		8 & 3.6649e-04 & $-$  & 1.8191e-04 & $-$  & 1.5034e-04 &  $-$ \\
		16 & 7.3066e-05 & 2.3265 & 3.9173e-05 & 2.2153 & 3.6042e-05 & 2.0605 \\
		32 & 1.5023e-05 & 2.2820 & 9.0355e-06 & 2.1162 & 8.8757e-06 & 2.0217 \\
		64 & 3.1943e-06 & 2.2336 & 2.1672e-06 & 2.0598 & 2.2074e-06 & 2.0075 \\
		128 & 7.0180e-07 & 2.1864 & 5.3059e-07 & 2.0302 & 5.5085e-07 & 2.0026 \\
		\hline
	\end{tabular}
\end{table}
\begin{table}[!htbp]
	\centering
	\caption{}{Results of comparison with $\mathfrak{p}=0.1$, $\mu=0.7$ and $J=150$}
	\label{tab:4}
	\begin{tabular}{lllllll}
		\hline
		\multicolumn{1}{c}{$\mathcal{N}$} &\multicolumn{2}{c}{$\text{Present~method}$ } & \multicolumn{2}{c}{$\text{Method~in~\cite{Golba}}$}& \multicolumn{2}{c}{$\text{Method~in~\cite{De}}$}\\
		$$ & $\mathfrak{L}_\infty$ &$\text{Rate}$& $\mathfrak{L}_\infty$ &$\text{Rate}$& $\mathfrak{L}_\infty$ &$\text{Rate}$\\
		\hline
		10 & 3.1579e-03 & $-$  & 5.8210e-03 & $-$  & 3.5000e-03 & $-$  \\
		20 & 1.2766e-03 & 1.3067 & 2.3040e-03 & 1.3372 & 1.4400e-03 & 1.3300 \\
		40 & 5.1746e-04 & 1.3028 & 9.0810e-04 & 1.3421 & 5.9000e-04 & 1.3150 \\
		80 & 2.0999e-04 & 1.3011 & 3.5720e-04 & 1.3461 & 2.4000e-04 & 1.3400 \\
		160 & 8.5257e-05 & 1.3004 & 1.4110e-04 & 1.3400 & 9.5000e-05 & 1.3600 \\
		320 & 3.4624e-05 & 1.3000 & 5.3870e-05 & 1.3892 & 3.8000e-05 & 1.3800 \\
		\hline
	\end{tabular}
\end{table}
\begin{table}[!htbp]
	\centering
	\caption{}{Results of comparison with $\mathfrak{p}=0.01$ and $J=100$}
	\label{tab:5}
	\begin{tabular}{lllllllll}
		\hline
		\multicolumn{1}{c}{$$} &\multicolumn{4}{c}{$\mu=0.9$ } & \multicolumn{4}{c}{$\mu=0.5$}\\
		\multicolumn{1}{c}{$\mathcal{N}$} &\multicolumn{2}{c}{$\text{Present~method}$}& \multicolumn{2}{c}{$\text{Method~in~\cite{Roul}}$}&\multicolumn{2}{c}{$\text{Present~method}$ } & \multicolumn{2}{c}{$\text{Method~in~\cite{Roul}}$}\\
		\hline
		256 & 1.9023e-04 & $-$  & 2.3339e-04 & $-$  & 9.9829e-06 & $-$  & 1.3091e-05 & $-$  \\
		512 & 8.8714e-05 & 1.1005 & 1.0896e-04 & 1.0989 & 3.5353e-06 & 1.4976 & 4.6540e-06 & 1.4920 \\
		1024 & 4.1380e-05 & 1.1002 & 5.0853e-05 & 1.0994 & 1.2518e-06 & 1.4978 & 1.6518e-06 & 1.4944 \\
		2048 & 1.9303e-05 & 1.1001 & 2.3729e-05 & 1.0997 & 4.4338e-07 & 1.4974 & 5.8557e-07 & 1.4961 \\
		4096 & 9.0050e-06 & 1.1000 & 1.1064e-05 & 1.1008 & 1.5731e-07 & 1.4949 & 2.0715e-07 & 1.4991 \\
		\hline
	\end{tabular}
\end{table}
\begin{table}[!htbp]
	\centering
	\caption{}{Results of comparison with $\mathfrak{p}=0.1$}
	\label{tab:6}
	\begin{tabular}{lllll}
		\hline
		\multicolumn{1}{c}{$$} &\multicolumn{2}{c}{$\mu=0.2$ } & \multicolumn{2}{c}{$\mu=0.7$}\\
		$(J, \mathcal{N})$& $\text{Present~method}$ &$\text{Method~in~\cite{Golba}}$& $\text{Present~method}$ &$\text{Method~in~\cite{Golba}}$\\
		\hline
		(4,4) & 8.0178e-04 & 3.1280e-02 & 1.0060e-02 & 2.0560e-02 \\
		(8,64) & 1.0281e-05 & 6.7910e-04 & 2.7454e-04 & 4.1590e-04 \\
		(16,1024) & 6.4780e-07 & 2.5030e-05 & 7.9848e-06 & 1.6720e-05 \\
		(8,8) & 1.9037e-04 & 1.3810e-02 & 4.1197e-03 & 1.0160e-02 \\
		(16,128) & 2.9444e-06 & 5.2460e-04 & 1.1409e-04 & 3.2180e-04 \\
		(32,2048) & 1.6358e-07 & 2.3070e-05 & 3.1896e-06 & 1.2730e-05 \\
		\hline
	\end{tabular}
\end{table}
\section{Conclusion}
This work proposed an efficient computational technique incorporating the Crank-Nicolson approach in time and exponential B-splines in the space for solving TFBSM. The developed algorithm is unconditionally stable with second order space accuracy and $2-\mu$ order time accuracy. We have presented numerical simulations to indicate the supremacy of the proposed scheme for solving pricing problems. The influence of the fractional orders on option pricings is highlighted through various plots. The numerical results produced via the exponential B-spline method are in good accordance with the methods available in the literature.
\section*{Acknowledgments}
The authors would like to extend their gratitude to anonymous editors and reviewers for their suggestions.

\section*{Financial disclosure}
Dr. Neetu Garg is supported by Faculty Research Grant by National Institute of Technology Calicut.
%


\begin{thebibliography}{6}
%
\bibitem{Black} 
Black F., Scholes M.: The pricing of options and corporate liabilities. J Polit. Econ. 81:637–654 (1973).
\bibitem{Merton}
Merton R.C.: Theory of rational option pricing. Bell. J. Econ. Manage. Sci. 4(1):141–183 (1973).
\bibitem{Podlubny}
 Podlubny I.: Fractional differential equations. Academic press, San Diego  (1999).
\bibitem{Wyss}
 Wyss W.: The fractional Black-Scholes equation. Fract. Calc. Appl. Anal. 3(3):51–-61 (2000).
\bibitem{Song}
 Song L., Wang W.: Solution of the fractional Black-Scholes option pricing model by finite difference method. Abstr. Appl. Anal. (1-2):194286 (2013).\url{doi:}
\bibitem{HZhang} 
Zhang H.,Liu F.,Turner I.,Yang Q.: Numerical solution  of the time fractional Black-Scholes model governing European options. Comput. Math. Appl. 71(1-4):1771--1783 (2016).
\bibitem{Cen} 
Cen Z., Huang J., Xu A., Le A.: Numerical approximation of a time-fractional Black-Scholes equation. Comput. Math. Appl. 75:2874–-2887  (2018).
\bibitem{Haq}
 Haq S., Hussain M.: Selection of shape parameter in radial basis functions for solution of time-fractional Black-Scholes models. Appl. Math. Comput. 335:248–-263 (2018).
\bibitem{Golba}
Golbabai A., Nikan O.: A computational method based on the moving least-squares approach for pricing double barrier options
in a time-fractional Black-Scholes model. Comput. Econ. 79:479--497 (2019).
\bibitem{Roul}
 Roul P.: A high accuracy numerical method and its convergence for time-fractional Black-Scholes equation governing European options. Appl. Numer. Math. 151:472--493 (2020).
     \bibitem{Ram1} 
 Ankur, Ram J., Naresh K.: Analysis and simulation of Korteweg-de Vries-Rosenau-regularised long-wave model via Galerkin finite element method. Comput. Math. Appl. 135:134--148 (2023).
 \bibitem{Ram2} 
 Ankur, Ram J., Akil N.: Conformal Finite Element Methods for Nonlinear Rosenau-Burgers-Biharmonic Models. arXiv preprint, https://arxiv.org/abs/2402.08926.
 \bibitem{Ram3} 
 Ankur, Ram J.: A new error estimates of finite element method for (2+ 1)-dimensional nonlinear advection-diffusion model. Appl. Numer. Math. 198:22--42 (2024).
   \bibitem{Ram4}  Ankur, Ram J.: New multiple analytic solitonary solutions and simulation of (2+1)-dimensional generalized Benjamin-Bona-Mahony-Burgers model, Nonlinear Dynamics, Pages 13297–13325,(2023).
      \bibitem{frac}
   Zhang X., Yang J., Zhao Y.: Numerical Solution of Time Fractional Black–Scholes Model Based on Legendre Wavelet Neural Network with Extreme Learning Machine. Fractal Fract. 6:401 (2022). 
\bibitem{Karatay}
Karatay I., Kale N., Bayramoglu S.R.: A new difference scheme for time fractional heat equations based on the Crank-Nicolson method. Frac. Calc. Appl. Anal. 16(4):892--910 (2013).
\bibitem{McCartin}
 McCartin B.J.: Theory of exponential splines. J. Approx. Theory 66(1):1--23 (1991).

 \bibitem{ravi}Ravi Kanth ASV, Neetu G.: An unconditionally stable algorithm for multiterm time fractional advection–diffusion equation with variable coefficients and convergence analysis. Numer Meth Part D E. 37(3) 1928--1945  (2020)
  \bibitem{ravi2}Ravi Kanth ASV, Neetu G.: A computational procedure and analysis for multi-term time-fractional Burgers-type equation. Numer Meth Part D E. 45(16) 9218-9232 (2022).

  \bibitem{De} 
  De Staelen R.H., Hendy A.S.: Numerically pricing double barrier options in a time-fractional Black-Scholes model. Comput. Math. Appl. 74:1166--1175 (2017).

\end{thebibliography}
\end{document}